\documentclass[twocolumn]{aastex63}

\usepackage{amssymb}
\usepackage{mathtools}

\def\msun{$M_{\odot}$}

\received{}
\revised{}
\accepted{\today}
\submitjournal{ApJ}

\shorttitle{Tycho SNR with LOFAR}
\shortauthors{Arias et al.}

\begin{document}

\title{Low-frequency radio absorption in Tycho's supernova remnant}

\correspondingauthor{Maria Arias}
\email{M.AriasdeSaavedraBenitez@uva.nl}

\author[0000-0002-0786-7307]{Maria Arias}
\affiliation{Anton Pannekoek Institute for Astronomy, University of Amsterdam, Science Park 904,1098 XH Amsterdam, The Netherlands}

\author{Jacco Vink}
\affiliation{Anton Pannekoek Institute for Astronomy, University of Amsterdam, Science Park 904,1098 XH Amsterdam, The Netherlands}
\affiliation{SRON, Netherlands Institute for Space Research, Sorbonnelaan 2, 3584 CA Utrecht}
\affiliation{GRAPPA, University of Amsterdam, Science Park 904, 1098 XH Amsterdam, The Netherlands}

\author{Ping Zhou}
\affiliation{Anton Pannekoek Institute for Astronomy, University of Amsterdam, Science Park 904,1098 XH Amsterdam, The Netherlands}

\author{Francesco de Gasperin}
\affiliation{Hamburger Sternwarte, Universit\"{a}t Hamburg, Gojenbergsweg 112, 21029, Hamburg, Germany}

\author{Martin~J. Hardcastle}
\affiliation{Centre for Astrophysics Research, School of Physics, Astronomy and Mathematics, University of Hertfordshire, College Lane, Hatfield AL10 9AB, UK}

\author{Tim~W. Shimwell}
\affiliation{ASTRON, Netherlands Institute for Radio Astronomy, Oude Hoogeveensedijk 4, Dwingeloo, 7991 PD, The Netherlands}
\affiliation{Leiden Observatory, Leiden University, PO Box 9513, 2300 RA Leiden, The Netherlands}



\begin{abstract}

Tycho's SNR is the remnant of the type Ia supernova explosion SN1572. In this work we present new low-frequency radio observations with the LOFAR
Low-Band and High-Band Antennae, centred at 58~MHz and 143~MHz, and with an angular resolution
of 41\arcsec\ and 6\arcsec\, respectively. We compare these maps to VLA maps at 327~MHz and 1420~MHz, and 
detect the effect of low-frequency absorption in some regions
of the remnant due to the presence of
free electrons along the line-of-sight. We investigate two origins for the low-frequency free-free absorption that we observe:
external absorption from foreground, and internal absorption from Tycho's unshocked ejecta.
The external absorption could be due to an ionised thin, diffuse cavity surrounding the SNR (although this cavity would need to be very
thin to comply with the neutral fraction required to explain the remnant's optical lines), or it could be due to an over-ionised molecular shell in the
vicinity of the remnant.
We note that possible ionising sources are the X-ray emission from Tycho, its cosmic rays, or radiation from Tycho's progenitor. 
For the internal absorption, we are limited by our understanding of the spectral
behaviour of the region at unabsorbed radio frequencies. However, the observations are suggestive of
free-free absorption from unshocked ejecta inside Tycho's reverse shock.

\end{abstract}



\section{Introduction}

Supernova remnants (SNRs) are the result of the interaction of a supernova explosion with its ambient medium.
The X-ray and radio-bright shell characteristic of young SNRs is composed of shocked ambient medium and
stellar ejecta. Internal to the reverse shock there can be some stellar ejecta that have yet to encounter the reverse shock \citep{mckee74}.
These ejecta were initially heated by the passage of the blast wave inside the star, but have since cooled due to adiabatic expansion.
Because this material is internal to a shell bright in X-rays and likely also in the UV, it can be photoionised.
Several hundreds of years after the supernova event, the remnant still retains some imprint of the explosion;
this is particularly the case for the unshocked ejecta.

SNRs have an effect on their surroundings, not only on the shocked ambient medium, but also on the
still to-be-shocked neighbourhood of the SNR. They are bright X-ray sources, as well as likely the sites of cosmic ray acceleration \citep{hillas05}.
Both the high-energy photons and the cosmic rays can deposit energy into the surroundings of the SNR;
for instance, heating and ionising nearby molecular clouds. Furthermore, during its lifetime and its
pre-SN stage, the progenitor star sculpts its ambient medium; for example, through stellar winds and ionising radiation.
The environment of the SNR is therefore a diagnostic of the star's pre-SN life, and of the SNR itself.

Tycho's SNR (SN 1572, G120.1+1.4, hereafter Tycho) is a young SNR, whose reverse shock might not have yet heated
all of the stellar ejecta from the explosion. It is the result of a Type Ia event, as evidenced
from the historical records of the light curve \citep{baade43}, and from the optical spectrum as recovered from
light echoes \citep{krause08b,rest08}.
From comparison of the X-ray spectra to hydrodynamical and spectral models, \cite{badenes06} concluded that the
scenario that best fit the data is one in which 1.3~\msun\ of material were ejected at the time of the explosion
into an ambient density of $\sim0.6-3$~cm$^{-3}$. There is evidence that the density is higher in the
north-east of the remnant, from H$\alpha$ \citep{ghavamian00}, molecular gas \citep{lee04, zhou16},
and dust observations \citep{williams13}. The work of \cite{woods17} placed strict upper limits on the temperature and luminosity of
Tycho's progenitor from the observed fraction of neutrals in the atomic gas, 
pointing to the merger of a double white dwarf binary as the most viable scenario for Tycho's SN explosion.
On the other hand, the molecular shell found in \cite{zhou16} is more consistent with a single-degenerate scenario.

The remnant has been studied extensively, including at wavelengths that probe the unshocked ejecta.
\cite{lopez15} observed it with \textit{NuStar}, but they did not not detect any  emission associated with the decay of
radioactive $^{44}$Ti, point-like or extended.
\cite{gomez12} observed it in the infrared with \textit{Herschel} and \textit{Spitzer}, and did not detect a cool dust component in the
innermost region of unshocked ejecta, although they did not specifically look for line emission from photoionised, cold material.
At low radio frequencies it has been observed with the Very Large Array (VLA) at 330~MHz \citep{katz-stone00}, and several times at
1.4 GHz \citep{reynoso97, katz-stone00, williams16}. It has also been observed at 660~MHz with the Westerbork Synthesis Radio
Telescope \cite[WSRT,][]{duin75}, and, at lower resolution, at 408~MHz as part of the Canadian Galactic Plane Survey  \cite[CGPS,][]{kothes06}.

In this paper we present new observations of Tycho with the LOw Frequency ARray \cite[LOFAR, ][]{vanhaarlem13}, both with the
instrument's High-Band Antenna (HBA, $120-168$~MHz) and the Low-Band Antenna (LBA, $40-75$~MHz). 
We compare these maps with higher frequency observations, and we detect localised free-free absorption from
free electrons along the line-of-sight, from foreground material, and possibly also
from material internal to the SNR reverse shock. 
We cannot use the measured absorption value to estimate how much mass there is in unshocked ejecta, although
our results suggest that if unshocked material is present, it is in a combination of relatively highly ionised, cold, and significantly clumped states.
The ionised ambient material could be either the diffuse cavity surrounding Tycho or its neighbouring molecular clouds.
Both scenarios have implications for the ionising source.

\section{Observations and data reduction}

\subsection{Observations}

We observed Tycho's SNR with LOFAR under project LC10\_011. The Low-Band Antenna (LBA) observations
were centred at RA=00:25:21.5, Dec=+64:08:26.9, with a time on-source of
10 hours. The data were taken on 2018/05/18, in the LBA-Outer configuration, using 8 bit sampling, 1 second integration,
and a frequency resolution of 64 channels per sub-band. 
The central frequency was 53.2 MHz, and the total bandwidth was 43.6 MHz.
A second beam was placed on calibrator 3C48 for the length of the observation.

For the High-Band Antenna (HBA) observations we made use of the possibility of co-observing with the LOFAR
Two Metre Sky Survey \cite[LoTSS,][]{shimwell17}. We identified the LoTSS pointing closest to Tycho, P007+64 (centred at
RA=00:30:40.8, Dec=+63:36:57.9), and requested that it be observed during LOFAR cycle 10 as part of LC10\_011. 
The observations were made with the standard LoTSS settings: 8 hours on-source, 48 MHz bandwidth, and an additional
10 minutes at the beginning and end of the observations to observe the calibrators (3C48 and 3C147, in this case). 

\subsection{Low-Band Antenna}

The LBA data were reduced with the LOFAR Low-Frequency Pipeline \citep{degasperin19}. The pipeline calibrates the calibrator
and transfers the solutions to the target, taking into account the main systematic effects in the LOFAR telescope,
such as clock drift, polarisation misalignment, ionospheric delay, Faraday rotation, ionospheric scintillation, beam shape, and bandpass.

Due to noise, we had to flag all the data at frequencies less than 40~MHz, as well as two LOFAR stations, CS013 and CS031. 
From the calibrator solutions we knew that there were very good ionospheric conditions during the observation, with
almost no Faraday rotation (the calibrator was observed for the full duration of the observation,
so we knew the ionosphere was good throughout). This allowed us to perform one round of self-calibration from our first image of the source,
rather than from a sky model made at a different frequency.

\begin{figure*}
\centering
\includegraphics[width=0.9\textwidth]{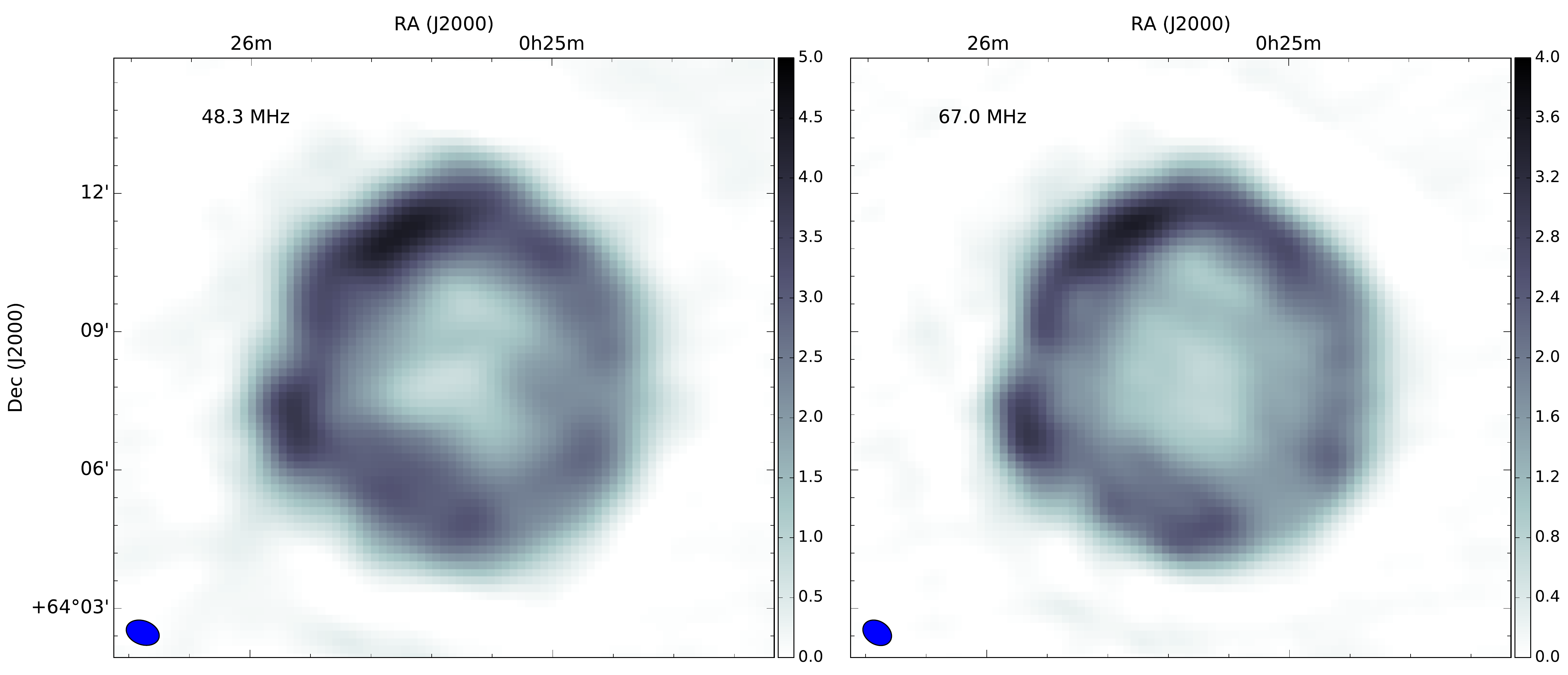}
\caption{Tycho SNR as observed with the LOFAR Low-Band Antenna (LBA). The LBA bandwidth was split to make these two images,
centred at 48.3~MHz (left) and 67.0~MHz (right), each 18~MHz wide. The elliptical beam size is 41\arcsec$\times31$\arcsec, with position angle $56^\mathrm{o}$,
and the pixel size is 10\arcsec for both maps. The local rms noise is 
0.03 Jy~bm$^{-1}$ for the 67.0~MHz map and 0.08 Jy~bm$^{-1}$ for the 48.3~MHz map. The flux density scale in both maps is in Jy~bm$^{-1}$.
\label{fig:lba_maps}}
\end{figure*}

The pipeline split the data into two frequency chunks, one centred at 48.3~MHz, and another centred at 67.0~MHz, which
were imaged separately. We imaged the data with \texttt{wsclean} \citep{offringa14}, which allows for multi-scale, multi-frequency 
deconvolution with w-projections, and for applying the LOFAR beam. The visibilities were weighted with a Briggs parameter of zero \citep{briggs95}.
In order to filter out large scale structure and in order to ensure common resolution among the maps, we used a $u-v$ range of
$30-5,000~\lambda$. The two \lq full-bandwidth' LBA images centred at 48.3~MHz and 67.0~MHz are shown in Fig. \ref{fig:lba_maps}.

In addition to the broadband maps, to search for spectral curvature, we made a series of narrow-band images, each 1.3~MHz wide, centred at 
40.1, 42.5, 44.8, 47.1, 49.5, 51.8, 54.2, 56.5, 58.9, 61.2, 63.5, 65.8, 66.9, 68.1, 70.5, 72.8, and 75.1 MHz.
These maps were also made with a common $u-v$ range of $30-5,000~\lambda$.

\subsection{High Band Antenna}

The HBA data were reduced in a direction-independent manner with the Pre-Facet Calibration Pipeline \citep{vanweeren16},
which obtains diagonal solutions towards the calibrator and then performs clock-TEC separation, 
which distinguishes between clock offsets and drifts, and signal delays due to the electron column density in the ionosphere,
and transfers the calibrator amplitudes and clock corrections to the data. 

\begin{figure}
\centering
\includegraphics[width=\columnwidth]{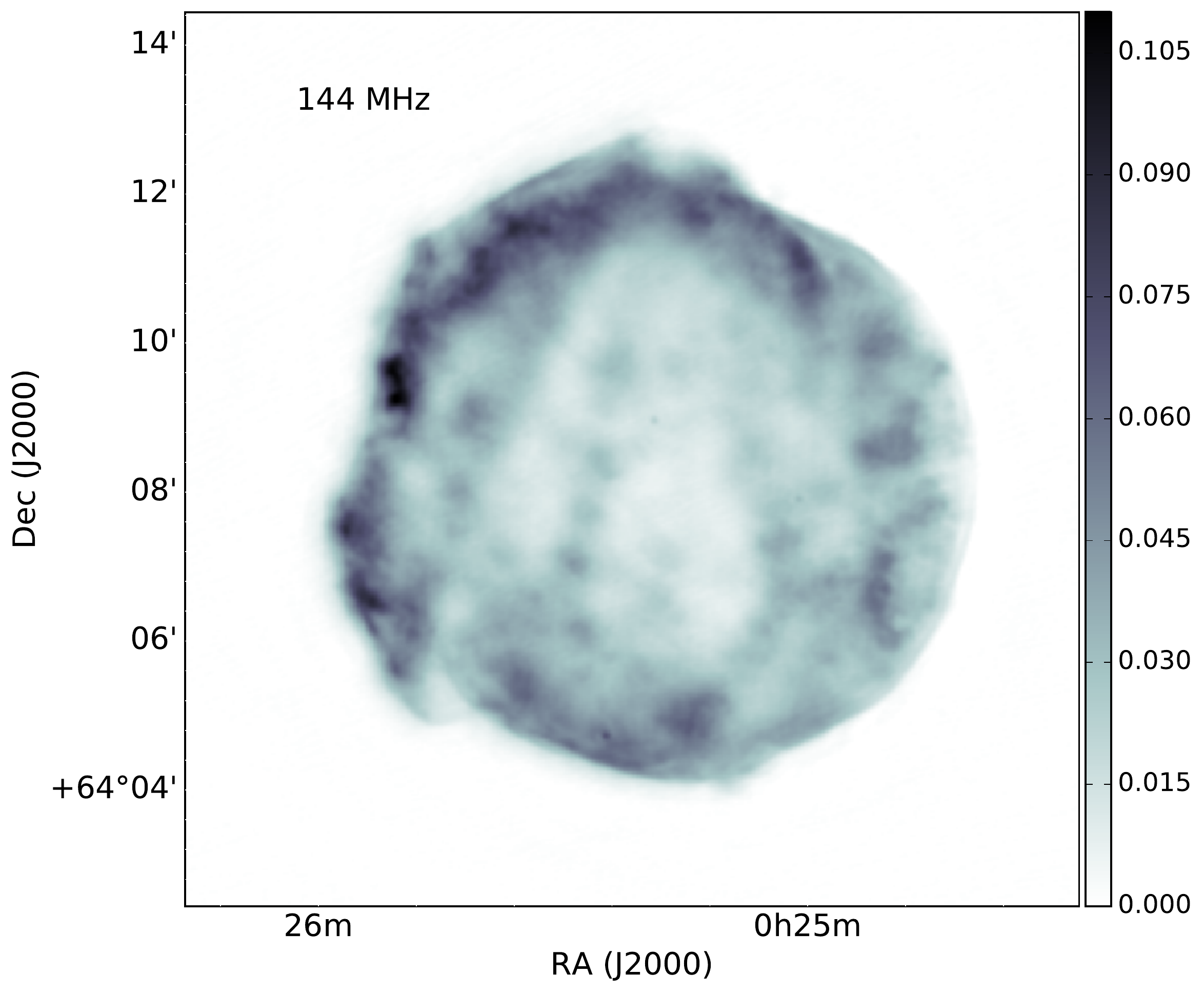}
\caption{Tycho SNR as observed with the LOFAR High-Band Antenna (HBA). The central frequency is 144~MHz, the bandwidth is 48 MHz,
the beam size is 6\arcsec, the pixel size is 1.5\arcsec, and the local rms noise is 1 mJy~bm$^{-1}$. The flux density scale is in Jy~bm$^{-1}$.
\label{fig:hba_map}}
\end{figure}

The calibrated data products were then imaged with the latest version of the ddf-pipeline\footnote{Version 2.2, \url{https://github.com/mhardcastle/ddf-pipeline/}} 
\citep{shimwell19,tasseinprep}, 
which is the method used for reducing data from the LoTSS. 
The pipeline carries out several iterations of direction-dependent self-calibration, using 
DDFacet for imaging \citep{tasse18} and KillMS for calibration \citep{tasse14a,tasse14b,smirnov15}. 
The resulting HBA image is shown in Fig. \ref{fig:hba_map}. The pipeline also produced three narrow-band images at 128, 144, and 160 MHz. The LOFAR
HBA in-band spectral index is unreliable, but in order to use these narrow-band maps in our analysis we bootstrapped the maps to the expected
flux densities of neighbouring sources in the field, from the HBA broadband map.

\subsection{Archival data}
\label{sec:archival}

We obtained the FITS files for the 327 MHz Very Large Array (VLA) observation of Tycho carried out in 1991-1993 \citep{katz-stone00},
as well as for the 1.4~GHz VLA observation carried out in 2013-2015 \citep{williams16}.
\cite{katz-stone00} note that their map is sensitive to scales between 8\arcsec\ and 30\arcmin, which corresponds
to $114-25,800 \lambda$s. The \cite{williams16} L-band map, combining the VLA A, B, C, and D configurations, is 
sensitive to scales between 1.3\arcsec\ and 16\arcmin\ ($212-15,800 \lambda$s).

The integrated flux density of the 1382~MHz map from \cite{williams16} is 41.7~Jy, and this is the value that we used for the 
analysis. However, if we directly measure the integrated flux density of the 327~MHz image, it is 121.8~Jy. This is 115\%
of the expected value for $S_\mathrm{1GHz}=56$~Jy and $\alpha=0.58$ \citep{green17}, and 117\% for $S_\mathrm{1GHz}=52.3$~Jy 
and $\alpha=0.63$, which are the best-fit values we find from a compilation of literature results (see discussion in section 
\ref{sec:flux}). We do not measure a level of background in the FITS image that accounts for this difference.
Unfortunately, \cite{katz-stone00} do not report the integrated flux density for their 327~MHz observation.

Our analysis relies on the localised deviation from power-law behaviour at low frequencies due to free-free absorption from ionised
material along the line-of-sight (we discuss the method in detail in section \ref{sec:method}). 
The 327~MHz and 1382~MHz maps provide the fit with the information about the spectral
behaviour of the source when no absorption is present. If we take the flux density at 327~MHz to be the 121.8~Jy that we measure directly
from the FITS file, we find it disproportionately affects the measured absorption, by setting an artificially high spectral index value for
any given pixel\footnote{The 121.8~Jy and 41.7~Jy values at 327~MHz and 1382~MHz correspond to a spectral index $\alpha_{327/1382}=0.74$,
much higher than the overall spectral index of the source.}, which then requires a much larger mass of absorbing material to account
for the flux densities at LOFAR frequencies. For this reason, we normalised the flux density of the 327~MHz map to 105.7~Jy, according to the
best-fit power law results for the compiled literature values as shown in section \ref{sec:flux}.

When comparing interferometric maps, it is important to take into account the scales probed by the different instruments.
When the emission is perfectly deconvolved, it
is possible to compare higher resolution maps with lower resolution maps by simply smoothing them to a common resolution.
However, the short-baselines $u-v$ coverage matters if interferometers do not probe the same scales, especially for
Galactic observations, for which the sources might be embedded in large-scale diffuse emission.

We summarise the  $u-v$ scales probed by the maps used in our analysis in Table \ref{tb:flux}.
Our LOFAR maps are sensitive to large angular scales, which might result in additional large-scale continuum emission that is
resolved out by the VLA maps. This would result in a spectral index steepening.
We note this issue as a possible source of error. 

\section{Results}

\subsection{Total flux density}
\label{sec:flux}

We report the total flux density of Tycho as seen with the LOFAR telescope LBA and HBA in Table \ref{tb:flux}.
We also include the values from the 327~MHz and 1382~MHz VLA observations \citep{katz-stone00, williams16} which we relied on for the analysis.

\begin{deluxetable}{c|cccc}[h]
\tablecaption{Flux densities of Tycho SNR \label{tab:fluxes}}
\tablehead{
\colhead{Freq} & \colhead{Flux density} & \colhead{Error} & \colhead{Year} & \colhead{$\lambda$ coverage}\\ 
\colhead{(MHz)} & \colhead{(Jy)} & \colhead{(Jy)} & \colhead{} \\
}
\startdata
48.3 & 334 & 33 & 2018 & $30-5,000~\lambda$ \\
67.0 & 275 & 27 & 2018 & $30-5,000~\lambda$ \\
144.6 & 163 & 16 & 2018 & $50-50,000~\lambda$ \\
\hline 
327 & 105.7& 10.5 &  1995 & $114-25,800 \lambda$ \\
1382 & 41.7 & 4.2 &  2013 & $212-15,800 \lambda$ \\
\enddata
\tablecomments{Observations at 327~MHz and 1382~MHz were taken with the
VLA and are described by \cite{katz-stone00} and \cite{williams16}, respectively.
See discussion in section \ref{sec:archival} for 327~MHz flux density.}
\label{tb:flux}
\end{deluxetable}

\begin{figure}
\centering
\includegraphics[width=\columnwidth]{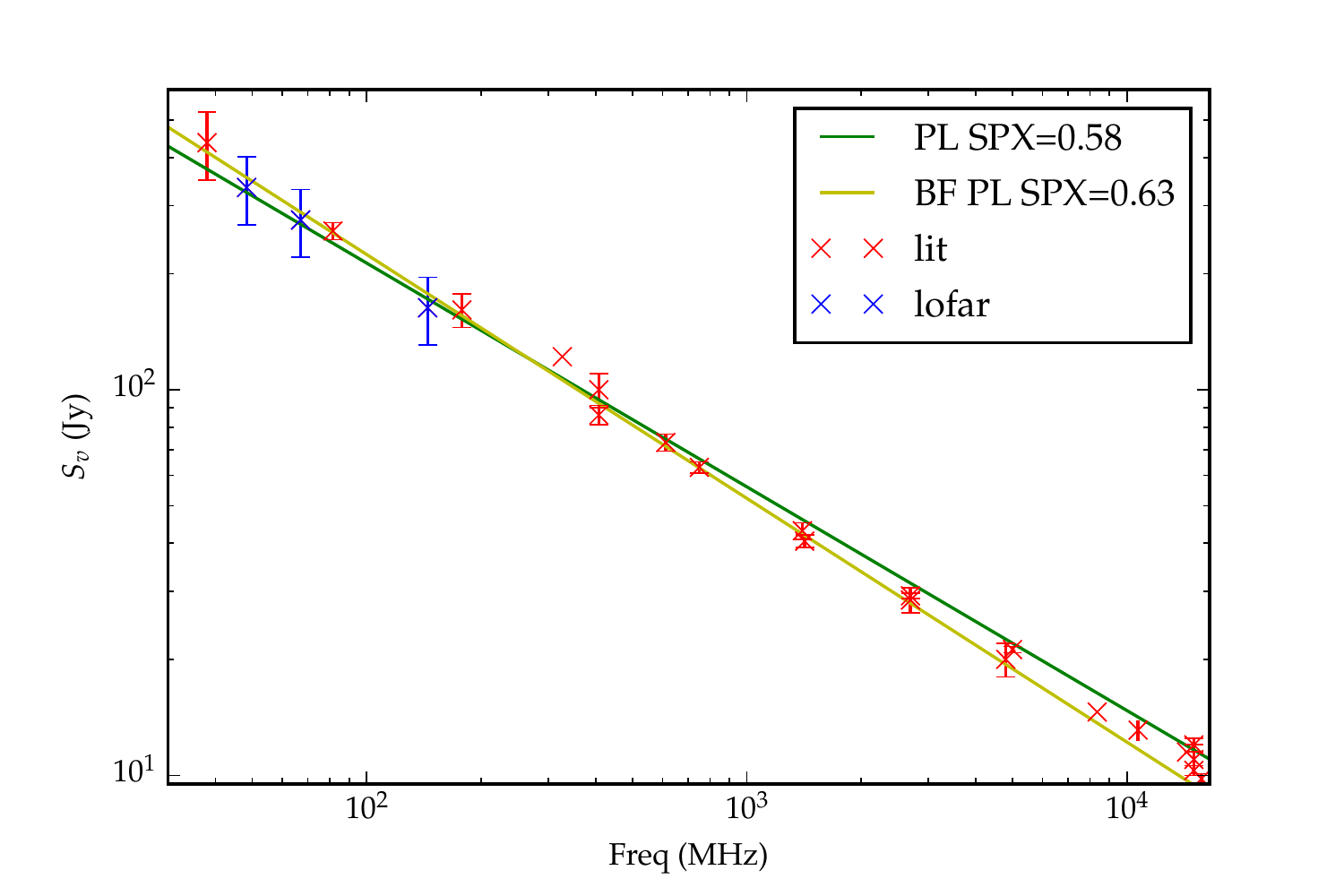}
\caption{Radio spectrum of Tycho, including measurements from this work (in blue). The green line corresponds
to the power-law spectral index (PL SPX) of 0.58 reported in \cite{green17}, and the yellow line is the best-fit (BF) power-law
spectral index from these data points. The literature (lit) values in red are
taken from: \cite{klein79}, \cite{green75}, \cite{hurley-walker09}, \cite{katz-stone00},
\cite{kothes06}, \cite{arnaud16}, \cite{gao11}, \cite{langston00}, \cite{williams66}, \cite{scott71}, \cite{artyukh69}, 
\cite{bennett63}, \cite{fanti74}, \cite{conway65}, \cite{kellermann69}, \cite{horton69}. 
\label{fig:radio_spectrum}}
\end{figure}

We compiled a series of radio flux densities in the literature, and plotted the LOFAR values alongside them
(Fig. \ref{fig:radio_spectrum}). 
Fitting a function of the form $S_\nu = S_\mathrm{1GHz} \left(\frac{\nu}{\mathrm{1GHz}}\right)^{-\alpha}$ gives 
a best-fit $S_\mathrm{1GHz}=52.3\pm2.0$~Jy and $\alpha=0.63\pm0.02$, whereas the value listed in the Green
SNRs catalogue is $S_\mathrm{1GHz}=56$~Jy and $\alpha=0.58$ \citep{green17}.

The systematic calibration errors in the LOFAR flux scale are of the order of 10\%, which dominates the uncertainties, rather than
the noise. For this reason we take 10\% errors when we report the integrated flux densities of Tycho in the broadband images
in Table \ref{tb:flux} and in Fig. \ref{fig:radio_spectrum}. However, the 10\% errors are on the total flux scale rather than the disagreement between in-band measurements. 
They are therefore an over-estimate for the purposes
of our analysis (our fits result in residuals that are much smaller than the error bars). 
The fact that we do not know the statistical errors of the flux densities presents an issue for the analysis.

\begin{figure}
\centering
\includegraphics[width=\columnwidth]{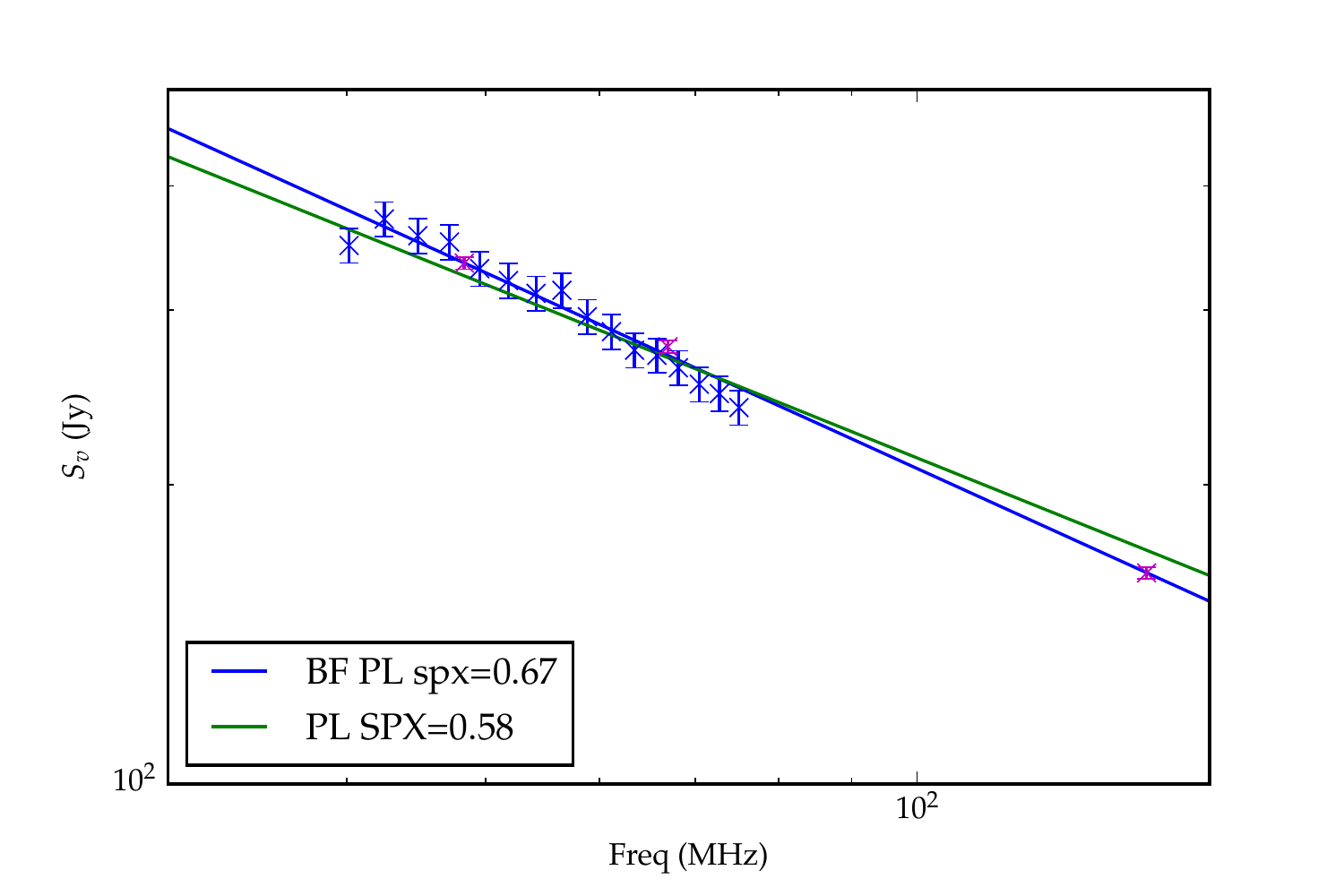}
\caption{Radio spectrum of Tycho at LOFAR frequencies. The magenta points correspond to the full bandwidth maps,
and the blue points correspond to the narrow band maps. 
 The green line corresponds
to the power-law spectral index (PL SPX) of 0.58 reported in \cite{green17}, and the blue line is the best-fit (BF) power-law
spectral index from the LOFAR data points.
The errors bars have been normalised so the reduced
$\chi^2$ of the best-fit power-law (in blue) is equal to 1, but we note that the uncertainties in the
LOFAR in-band have not been systematically analysed and can be unreliable. 
Our measurements agree with earlier reports that the radio spectrum of Tycho steepens at low radio frequencies.
\label{fig:lofar_spectrum}}
\end{figure}

In order to solve this problem,  
we artificially shrank the error bars of the LOFAR images (see Fig. \ref{fig:lofar_spectrum}) until
the reduced $\chi^2$ of the best-fit power-law for these points was 1. This provides us with a more meaningful estimate
of the errors in our pixel-by-pixel analysis.

The flux densities of the LBA narrow-band maps are plotted in Fig. \ref{fig:lofar_spectrum}. 
If we only consider the LOFAR LBA and HBA results, we measure a steeper spectral index than when
we take into account measurements at higher frequencies ($\alpha=0.67$ instead of $\alpha=0.58$ or $\alpha=0.63$).
The best-fit value of $\alpha$ for the LOFAR points ($\alpha=0.63$) results in a $\Delta\chi^2=23.7$ improvement over the fixed $\alpha=0.58$ scenario,
for one additional degree of freedom.

\subsection{Model parameters: external absorption}
\label{sec:method}
A synchrotron source with spectrum $S_\nu \propto \nu^{-\alpha}$ that is subject to free-free absorption
from cold, ionised, ISM material along the line of sight results in the following radio spectrum:
\begin{equation}
S_\nu = S_0 \left( \frac{\nu}{\nu_0} \right)^{-\alpha} \, e^{-\tau_{\nu, \mathrm{ISM}}},
\label{fitting}
\end{equation}
where \citep{rybicky79}:
\begin{equation}
\tau_\nu = 3.014 \times 10^{4} \, Z \,\left( \frac{T}{\rm{K}} \right)^{-3/2} \left( \frac{\nu}{\rm{MHz}} \right)^{-2} \left( \frac{{EM}}{\rm{pc \,cm}^{-6}} \right) g_{\mathrm{ff}},
\label{ff_tau}
\end{equation}
$Ze$ is the charge of the free-free absorbing ions, $T$ is the temperature of the plasma, $EM\equiv \int_{0}^{s} n_\mathrm{e}^2 ds'$ is the emission measure,
$n_\mathrm{e}$ is the number density of electrons,
and $g_{\mathrm{ff}}$ is a Gaunt factor, given by
\begin{equation}
g_{\mathrm{ff}} = 
\begin{cases}
\ln\left[49.55 \, Z^{-1} \left(\frac{\nu}{\rm{MHz}}\right)^{-1} \right] + 1.5 \ln \frac{T}{\mathrm{K}}  \\  \\  1 & \hspace{-4cm} \text{for} \,\,\, \frac{\nu}{\rm{MHz}}>>\left(\frac{T}{\mathrm{K}}\right)^{3/2}.
\end{cases}
\end{equation}

\begin{figure*}
\centering
\includegraphics[width=\textwidth]{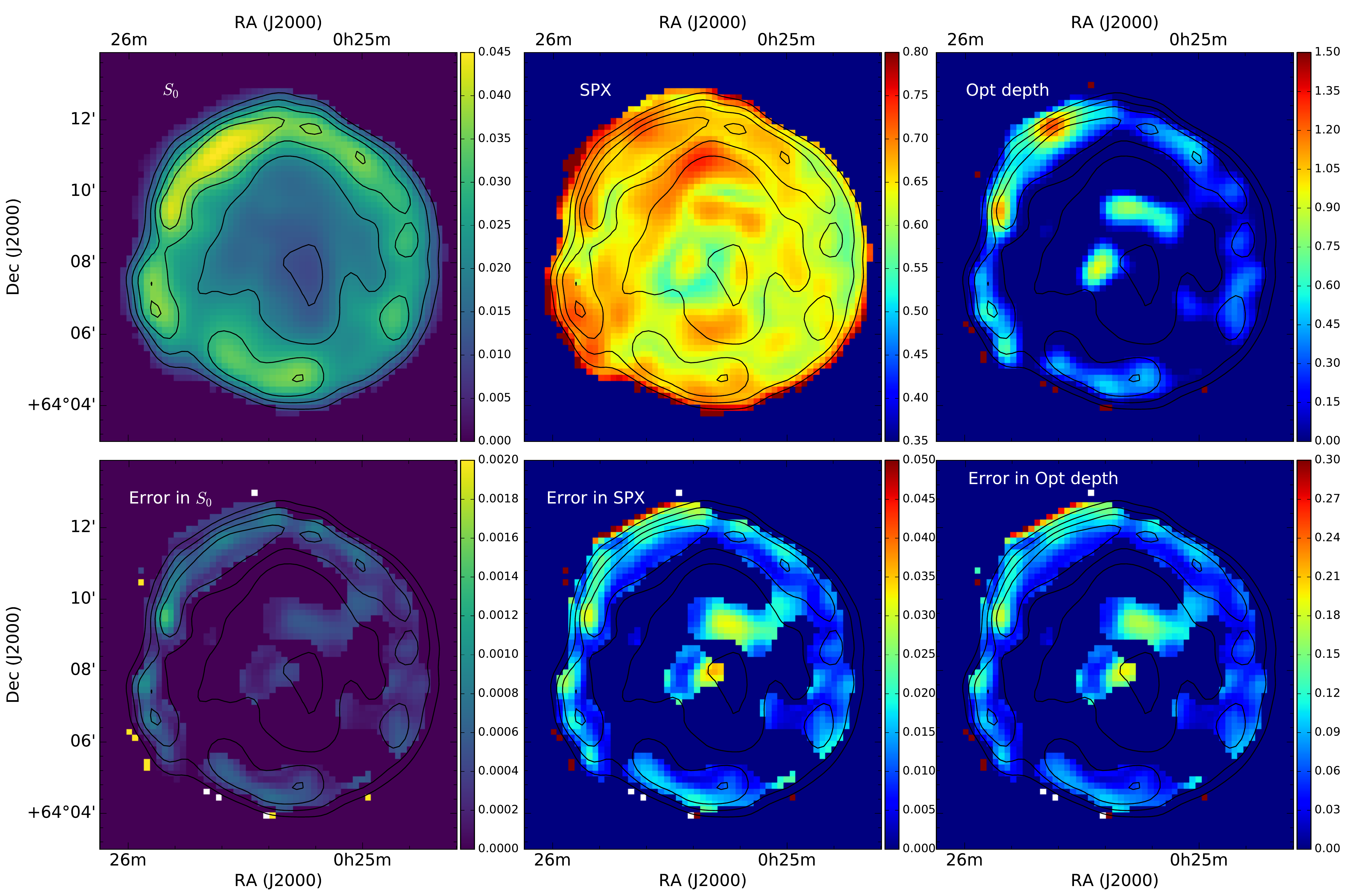}
\caption{Results of fitting equation \ref{fitting} to the maps. For each pixel we fitted for amplitude $S_0$, the spectral index $\alpha$,
and the optical depth for the ISM material at 40~MHz $\tau_{40, \mathrm{ISM}}$. The units of the $S_0$ map on the left are Jy~bm$^{-1}$.
The errors are the diagonal term of the covariance
matrix corresponding to each parameter.
\label{fig:results}}
\end{figure*}

We convolved all the images to a resolution of 41\arcsec, and performed a pixel-by-pixel fit (with a pixel size of 10\arcsec) to equation \ref{fitting}. The
results are plotted in Fig. \ref{fig:results}. For each pixel, we fitted for an amplitude $S_0$, the spectral index $\alpha$,
and the optical depth for the ISM material at 40~MHz $\tau_{40, \mathrm{ISM}}$. As errors, we plot the diagonal term of the covariance
matrix corresponding to each parameter.

\begin{figure*}
\centering
\includegraphics[width=0.9\textwidth]{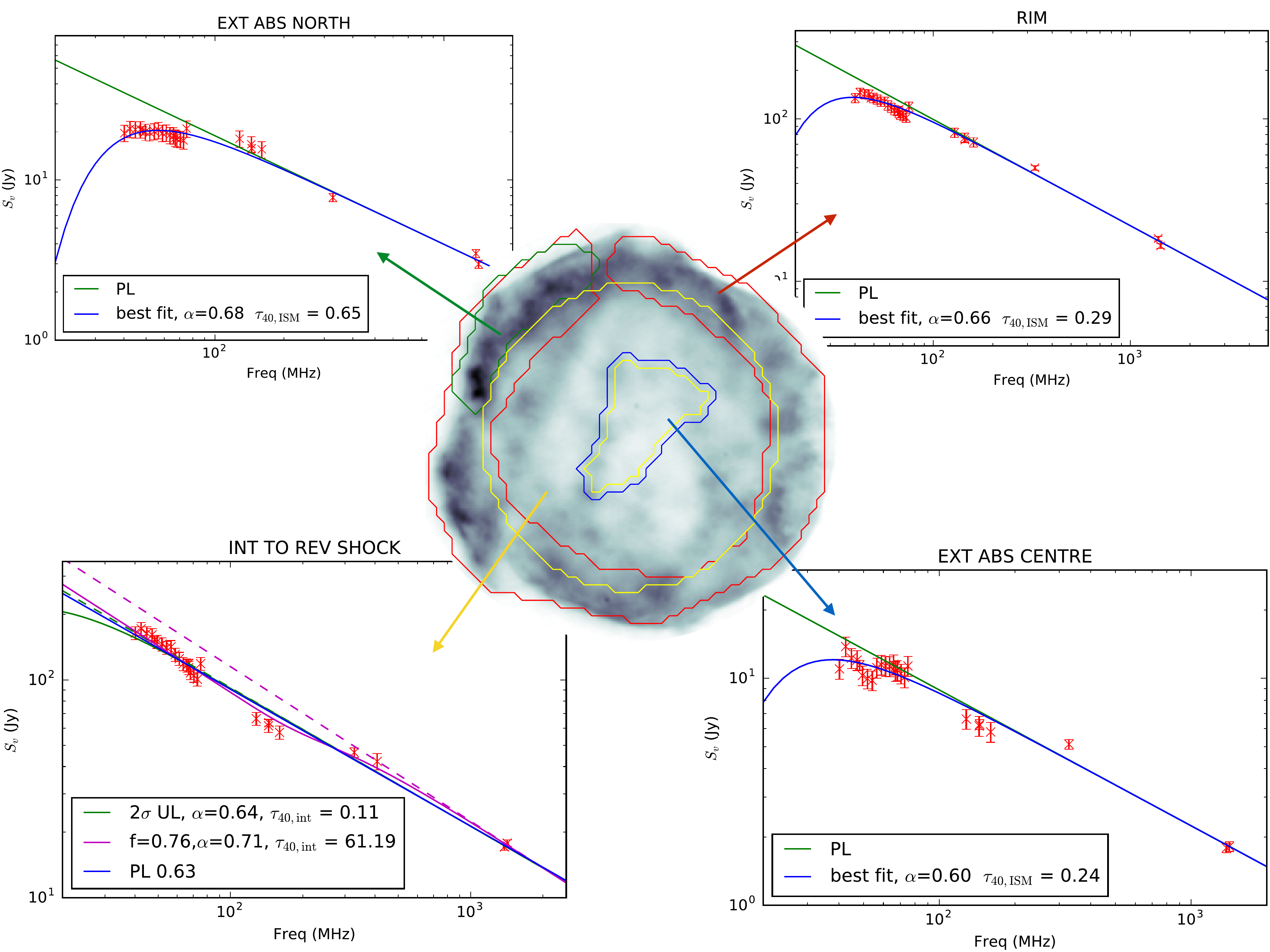}
\caption{HBA map with overlaid regions of analysis. The values of $f$, $\tau_{40}$ and $\alpha$ are unitless. For all regions, the errors were rescaled in such a way that the best-fit
power law has a reduced $\chi^2$ of 1. The top plots and the bottom-right plot (corresponding to the green, red, and blue regions as overlaid on Tycho)
are fitted including external absorption (in blue, the
best-fit unabsorbed power-law is in green), and in all cases including the absorption term improves the fit: with a $\Delta\chi^2=16$ for
\lq EXT ABS NORTH', a $\Delta\chi^2=4$ for \lq EXT ABS CENTRE', and a  $\Delta\chi^2=10.5$ for \lq RIM' (in all cases, for an additional
degree of freedom). The bottom-left plot corresponds to the region of possible internal absorption.
The mask of the reverse shock radius is plotted in yellow over the map of Tycho. In the legends, \lq UL' stands for \lq upper limit' and 
\lq PL' stands for \lq power-law'.
\label{fig:hba_reg}}
\end{figure*}

\begin{figure*}
\centering
\includegraphics[width=\textwidth]{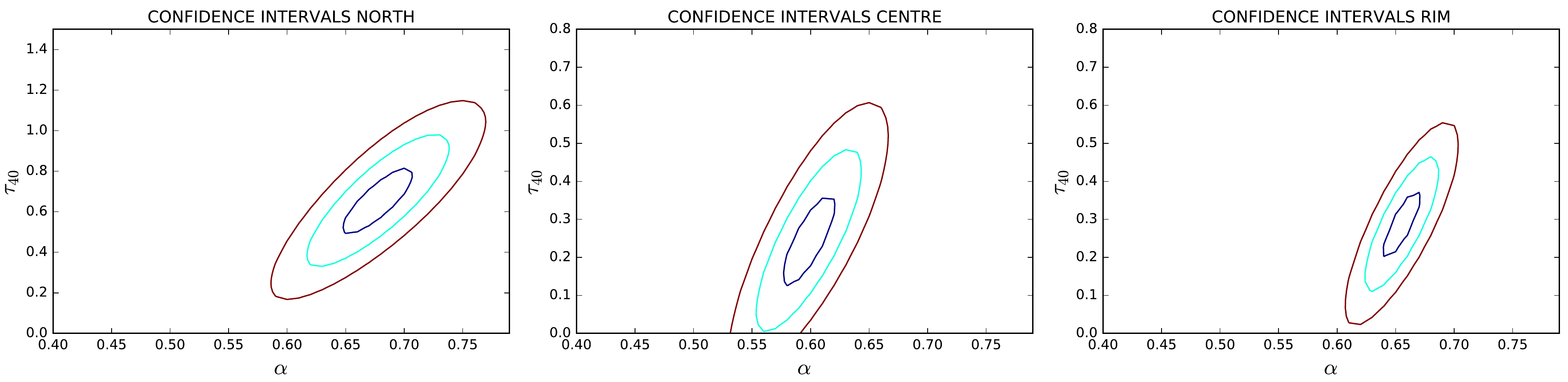}
\caption{Contour plots for the three regions showing external absorption in Fig. \ref{fig:hba_reg}: north (shown in red over Tycho in Fig. \ref{fig:hba_reg}), centre (in blue),
and rim (in red). Plotted are the $1\sigma$, $2\sigma$, and $3\sigma$ confidence intervals for the parameters $\alpha$ and $\tau_{40,\mathrm{ext}}$ for each of the regions.
Only for one region, centre, $\tau_{40,\mathrm{ext}}=0$ (no absorption) is not excluded at the $3\sigma$. For the two other regions, in particular for the northern region
that we base our analysis on, we require the presence of absorption along the line-of-sight at the $3\sigma$ level.
\label{fig:conf_intervals}}
\end{figure*}

We also show the fit results for three integrated regions that show external absorption (see Fig. \ref{fig:results}, right panel): the region towards the north-east, 
the absorbed region in the centre, and the whole rim of the SNR. These regions are labeled in Fig. \ref{fig:hba_reg}, and their spectral energy distribution (SED)
along with the best-fit
results are shown. The parameters $\alpha$ and $\tau_{40,\mathrm{ext}}$ are correlated (see contour plots in Fig. \ref{fig:conf_intervals}), but for two of the
three regions we require absorption at the $3\sigma$ level or higher.

\subsection{Model parameters: internal absorption}

A synchrotron source that is subject to internal free-free absorption
from its cold, ionised, unshocked ejecta will have a dimming factor that goes as $(f + (1-f) e^{-\tau_\nu})$, where $f$ is
the fraction of the synchrotron emission that is produced by the front side of the shell and, therefore, cannot be absorbed by its
internal material. This factor multiplies equation \ref{fitting} resulting in the following radio spectrum:
\begin{equation}
S_\nu = S_0 \left( \frac{\nu}{\nu_0} \right)^{-\alpha} (f + (1-f)e ^{-\tau_{\nu, \mathrm{int}}}) \, e^{-\tau_{\nu, \mathrm{ISM}}}.
\label{fitting_all}
\end{equation}
Internal free-free absorption can only occur in the region inside the projected reverse shock, since there cannot be unshocked absorbing material outside the reverse shock. 

\cite{warren05} found the reverse shock in Tycho's SNR to have a radius of 183\arcsec\ and centre
RA=0:25:19.40, Dec=+64:08:13.98, from principal component analysis of X-ray data. We measured the flux density for
each image for the region internal to the reverse shock, with the aim to look for internal
absorption. 
We do not find any external absorption in the region internal to the reverse shock, save for two clumps in the center
of the SNR (Fig. \ref{fig:results}), and so, to simplify our fit, we
we removed the area of absorption in the centre (the blue region in Fig. \ref{fig:hba_reg}) from our area of internal absorption
(the yellow region in Fig. \ref{fig:hba_reg}), and just fitted for an amplitude, the parameter $f$, and an internal optical depth:
\begin{equation}
S_\nu = S_0 \left( \frac{\nu}{\nu_0} \right)^{-\alpha} (f + (1-f)e ^{-\tau_{\nu, \mathrm{int}}}).
\label{fitting_all}
\end{equation}

As described in section \ref{sec:flux}, we rescaled the error bars in such a way that
the reduced $\chi^2$ of the best-fit power-law ($S_\nu = S_0 \left( \frac{\nu}{\nu_0} \right)^{-\alpha}$, with no absorbing component) was 1. The best-fit power-law
for this region corresponds to $\alpha=0.63$. From here, we compared how including an 
internal absorbing component improved the fit.

Setting $f=0.5$ (that is, fixing the synchrotron emission such that half comes from the back and half comes from the front
of the shell) gives a best-fit $\alpha=0.63$, $\tau_{40,\mathrm{int}}=3\times10^{-8}$.
This means that the best-fit value for internal absorption with $f=0.5$ corresponds to no internal absorption.
Setting $f=0.5$, 
for $\tau_{40,\mathrm{int}}=0.11$ we obtained a $\Delta \chi^2=4$ (with respect to the best-fit result). 
We take this to be the $2\sigma$ upper limit estimate on the 
internal optical depth, and so in the internal emission measure $EM_\mathrm{int}$ (for $T=100$, $Z=3$).

Alternatively, if we fit a region that shows internal absorption with a power-law, the spectral
index flattens due to the presence of absorption. \cite{katz-stone00} found
$\alpha=0.71$, rather than $\alpha=0.63$ for this region, from a 330~MHz to 1.4~GHz spectral index
study. In fact, the higher frequency data points, where no absorption is present, should be the ones
that determine the spectral index. At low frequencies the original spectral index should be recovered, but with the amplitude
dimmed by a factor of $f$. Hence, we fixed the spectral index to $\alpha=0.71$ and fitted for the remaining parameters.
This results in a very high value of the optical depth, $\tau_{40,\mathrm{int}}=61$.

\begin{deluxetable}{c|ccccc}[h]
\tablecaption{Fits to region internal to the reverse shock}
\tablehead{
\colhead{Fit} & \colhead{$\alpha$} & \colhead{$f$} & \colhead{$\tau_{40, \mathrm{int}}$} & \colhead{red $\chi^2$} & \colhead{$\Delta \chi^2$} \\ 
}
\startdata
PL & 0.63 & $-$ & $-$ & 1.0 & - \\
best-fit int abs & 0.63 & 0.5* & $3\times10^{-8}$ & 1.1 & 0 \\
UL in int abs & 0.64 & 0.5* & 0.11 & 1.3 & 4 \\
Fixed $\alpha$ & 0.71* & 0.76 & 61.1 & 0.4 & 16 \\
\enddata
\tablecomments{The best-fit emission measure $EM$ assumes $T=100$~K and $Z=3$. Parameterised, it corresponds to
$EM = EM_\mathrm{table} \mathrm{\, pc \, cm}^{-6} \left( \frac{g_\mathrm{ff}(T=100, Z=3)}{g_\mathrm{ff}(T/100 \,\mathrm{K}, Z/3)} \right)
\times \left( \frac{Z}{3} \right) \left( \frac{T}{100 \,\mathrm{K}} \right)^{-3/2}$. The reduced $\chi^2$ to the
power-law fit is 1 by definition. Values indicated with * are fixed, not fitted for. The $\Delta \chi^2$ for the \lq Fixed $\alpha$' model
is with respect to the power-law model \lq PL', corresponding to 2 additional degrees of freedom. The upper limit \lq UL' was 
derived as discussed in the text.
}
\label{tb:fits_int}
\end{deluxetable}

The results of our fits \cite[power-law, internal absorption, $2\sigma$ upper limit in internal absorption, and $\alpha$ fixed to the value
given by][]{katz-stone00} are tabulated in Table \ref{tb:fits_int}. 
We also plotted the results for the power-law fit (in blue), the upper limit to the $EM$ (for $T=100$, $Z=3$, in green)
and the fixed $\alpha$ (in magenta; dashed lines indicate the unabsorbed flux density)
in Fig. \ref{fig:hba_reg}, bottom-left corner. Here we show the rescaled errors
rather than the original error bars.

From Table \ref{tb:fits_int}, fixing $\alpha=0.71$ and adding an absorbing component does seem to significantly improve the fit
(the fact that the reduced $\chi^2$ is equal to 0.4 would normally suggest overfitting, but in this case the reduced $\chi^2$ of the
power-law fit was artificially set to 1). The
required emission measure is unphysical (see discussion in section \ref{ush_mass}), but it is very sensitive to the choice
of $\alpha$ and $f$. We cannot confidently
claim a detection of unshocked ejecta in Tycho's SNR because of our limited knowledge of the errors in the flux densities, and because of the degeneracy of the parameters,
but our data are suggestive that there is indeed some unshocked material inside Tycho's reverse shock\footnote{
In the conference Supernova Remnants: An Odyssey in Space after Stellar Death II (Chania, Greece, June 2019)
we presented preliminary results
of a very high $EM$ detection from Tycho's unshocked ejecta (\url{http://snr2019.astro.noa.gr/wp-content/uploads/2019/08/D3-0940-Arias.pdf}). 
This was due to us not noticing at first that the 330~MHz map had a very high flux density value,
which steepened the best-fit spectral index, and thus the required amount of absorbing material.}.

In order to better estimate the $EM$ due to internal absorption we need more high-frequency data points in the  few GHz range that can unambiguously
determine the unabsorbed flux density and spectral index for this region. Additional observations in the few-hundred MHz range
would help better model the curvature due to the free-free absorption, and, if it were ever possible, observations at even lower frequencies
would further discriminate between the different models.
In this work we are relying on only the points at 327~MHz and 1382~MHz for
information about the unabsorbed flux density and spectrum,
and the 327~MHz map was rescaled (see discussion in section \ref{sec:archival}). Moreover, the behaviour of the LOFAR
in-band seems to be pushing the data point in a steeper spectral index direction.
For this reason, observations that increase the leverage arm in frequency would allow us to better constrain the amount of $EM$ due to
unshocked material.

Having said that,
the integrated flux densities as measured by LOFAR
are in line with what we expect from the literature. 
There are some regions where the maps can have artefacts,
but the flux densities that we are considering in this section are taken from the yellow region in 
 Fig. \ref{fig:hba_reg}, which is much larger than the resolution
of any given map. 
Moreover, the LBA and the HBA data both show the effect of absorption, even though
the two LOFAR antennas are effectively different instruments, and the data were reduced with two independent pipelines.

\section{Discussion}

\subsection{Spectral index}

\cite{katz-stone00} carried out a study of Tycho's spectral index at low radio frequencies (330~MHz and 1.5~GHz), and found that Tycho 
has localised spectral variations with regions as flat as $\alpha=0.44$ and as steep as $\alpha=0.72$. Our best-fit spectral index map 
(middle panel in Fig. \ref{fig:results}) shows values within this range, and, in a few cases, slightly higher values, $\alpha \lesssim 0.8$.

\cite{duin75} reported a significant steepening of the spectrum near the centre of the SNR and suggested that particles
near the boundary might be accelerated with a flatter spectrum, but \cite{klein79} did not find steepening in their observations at 10 GHz.
We do not find a steepening coincident with the centre of the remnant, but rather we find the
spectrum of the western and north-western region of the remnant to be 
steeper than the rest. 

The question of whether Tycho has a curved spectrum has been discussed in the literature. 
\cite{roger73} modelled Tycho's integrated radio spectrum with two power-law components (which results in a locally concave spectrum), 
\cite{reynolds92} modelled it with a non-linear shock model of first-order Fermi acceleration and found agreement with a concave-up
synchrotron spectrum, whereas \cite{vinyaikin87} found that a single power-law can describe the radio spectrum at these frequencies. 
As we discussed in section \ref{sec:flux}, the LOFAR data points do show a steeper spectral behaviour than expected, although the
in-band response of the LOFAR LBA has not been systematically analysed, and is not yet reliable. 

\subsection{External absorption}
\label{sec:ext_abs}

\begin{figure*}
\centering
\includegraphics[width=\textwidth]{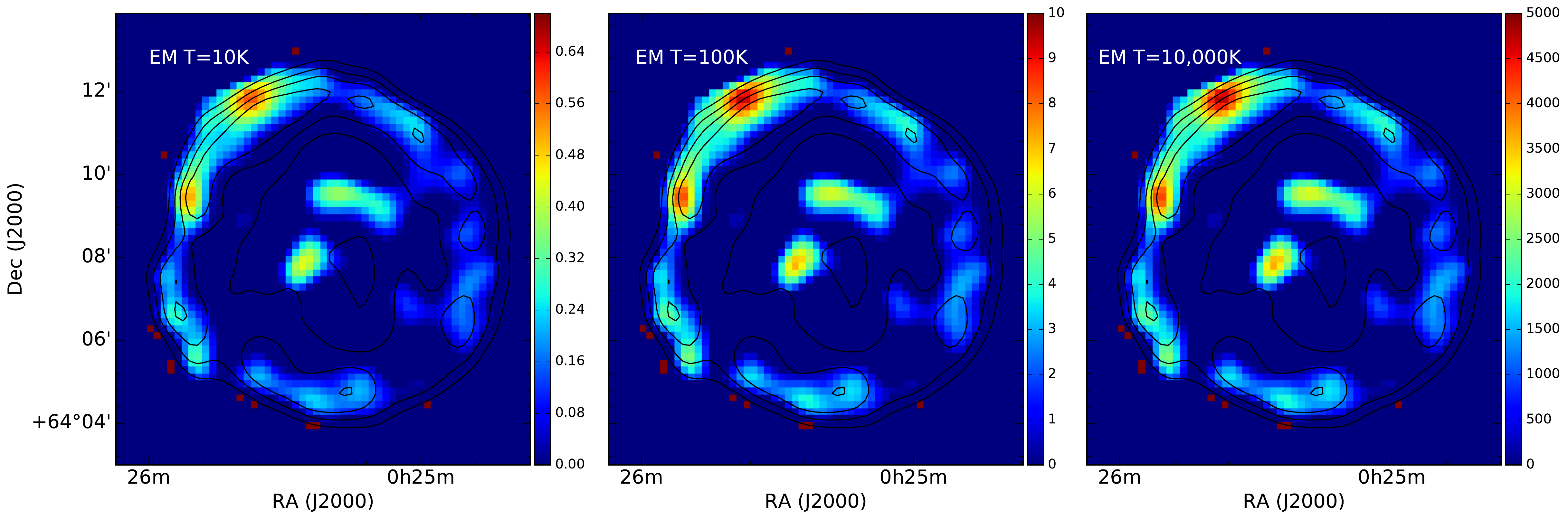}
\caption{Maps of external emission measure $EM_\mathrm{ISM}$ made from the measured optical depth $\tau_{40,\mathrm{ISM}}$ 
(right hand side map in  in Fig. \ref{fig:results}) combined with equation \ref{ff_tau}, assuming $Z=1$.
We plot the results for three temperatures, 
10~K, 100~K, and 10,000~K, relevant for our discussions of molecular clouds, the diffuse, infrared-emitting medium around Tycho,
and the ISM warm ionised gas, respectively.
The units of
$EM_{\mathrm{ISM}}$ are $\mathrm{\, pc \, cm}^{-6}$.
\label{fig:Ems}}
\end{figure*}

In order to convert the value of optical depth in Fig. \ref{fig:results} into a quantity that allows us to derive physical properties of the
gas we use equation \ref{ff_tau}, from which we obtain an emission measure value, $EM_{\mathrm{ISM}}$.
The emission measure depends on the temperature and ionisation state of the plasma. The ISM has a wide range of temperatures,
from $\sim10$~K in molecular clouds to $\sim10,000$~K in the warm ionised medium \citep{draine11}. We therefore
provide three emission measure maps in Fig. \ref{fig:Ems}, assuming $T=10$~K, $T=100$~K, and $T=10,000$~K, to aid our discussion
in the current section. Since the ISM is primarily composed of hydrogen, for all three maps we assume $Z=1$.

The region to the north-east with the high emission measure value (the region in green in Fig. \ref{fig:hba_reg}) 
seems to match the position of a molecular cloud
found in \cite{lee04} and \cite{zhou16}, seen most clearly in Fig. 1 of the latter paper at velocities between $-62$~km~s$^{-1}$ and $-66$~km~s$^{-1}$.
At these velocities there are also multiple structures that coincide in position with the rim of the source, which our fit also identifies
as having free-free absorption. The region in the north-east of the remnant where we find the
highest values of the $EM_{\mathrm{ISM}}$ also coincides with the region of high H I absorption seen in \cite{reynoso99b}. 
The region in the centre of Tycho has some morphological coincidence with the molecular structure seen at $-56$~km~s$^{-1}$ in
\cite{zhou16}, although the similarity is not striking, and there
does not seem to be any associated neutral hydrogen structure.
Our method traces ionised
material, which one does not expect in molecular clouds but could be present at their outer boundary, so it is not necessary that 
our measured $EM_{\mathrm{ISM}}$ matches the structure of molecular/neutral material in detail. 

The scale and distance of the ionised features are not straightforward from these observations. 
Tycho is the background synchrotron source, so the ionised material must be in front
of it, but in principle it could be local to Tycho, unrelated ISM, or a combination of the two (although it would be a big coincidence if
one of the two did not have a dominant effect).

\begin{figure*}
\centering
\includegraphics[width=\textwidth]{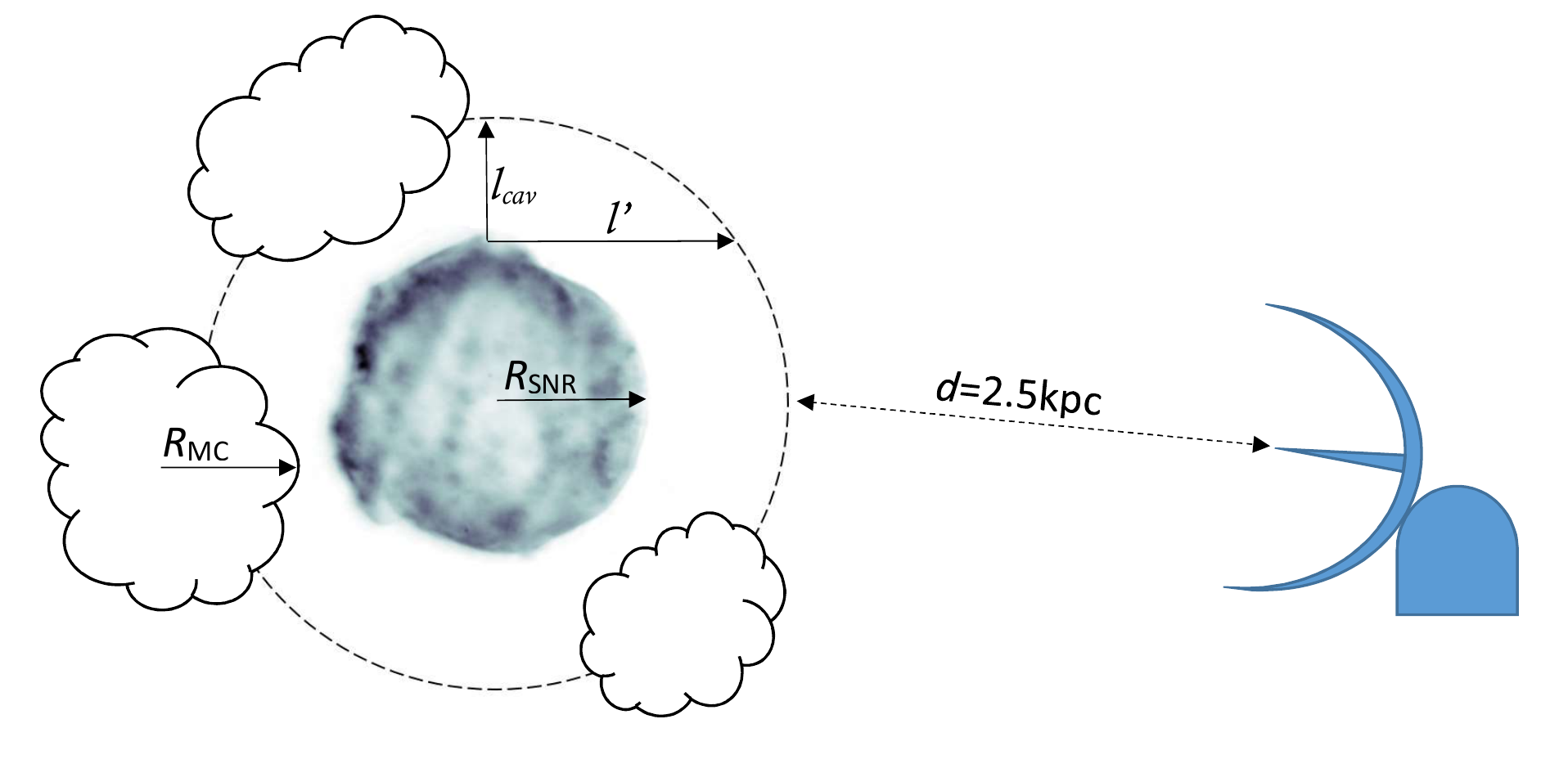}
    \caption{Cartoon showing the geometry assumed for the discussion in section \ref{sec:ext_abs}.
    Tycho is surrounded by a diffuse cavity of length $l_\mathrm{cav}$, and the molecular clouds are in a ring-like
    shape around it. 
\label{fig:cartoon}}
\end{figure*}

We know from \cite{zhou16} that Tycho is likely inside an expanding wind bubble that is sweeping up molecular material. 
We depict the structure we assume for our analysis in a cartoon in Fig. \ref{fig:cartoon}. The remnant is surrounded by,
but its shock is still not interacting with, molecular clouds. This means that there is a cavity of thickness $l$ (and radius $R_\mathrm{SNR}+l$)
of low-density material \cite[$\rho=0.1-0.2$~cm$^{-3}$,][]{williams13}, surrounded by dense molecular material with
an average density of $10^2-10^3$~cm$^{-3}$ \citep{zhou16}.

We will consider three possibilities: (1) that the ionised material we see in Fig. \ref{fig:results}, right-hand side, is due to ionised material
along the line-of-sight, unrelated to Tycho; (2) that it is the low-density cavity material that is ionised; and (3) that the molecular clouds
are responsible for the free-free absorption. 
In section \ref{sec:source} we briefly mention possible ionising sources.

\vspace{0.2cm}
\noindent
\textbf{1. Ionised ISM}

\cite{hwang02} tabulated the $N_\mathrm{H}$ as measured from \textit{Chandra} data, towards Tycho, and found values ranging from
$N_\mathrm{H}=(5.3-7.5)\times10^{21}$~cm$^{-2}$, depending on the model employed. 

For the region in green in Fig. \ref{fig:hba_reg} the optical depth at 40~MHz is $\tau_{40, \mathrm{ISM}}=0.65$,
which corresponds to an 
emission measure of $EM=0.30$~pc~cm$^{-6}$ for $T=10$~K, and $EM=2469$~pc~cm$^{-6}$ for $T=10,000$~K. Since $EM = n_\mathrm{e}^2 l$,
$N_\mathrm{H} = n_\mathrm{H} l$, and $n_\mathrm{e} = \chi_\mathrm{e} n_\mathrm{H}$ (where $\chi_\mathrm{e}$ is the ionisation fraction,
$0\leq\chi_\mathrm{e}\leq1$),
then, using $l=d$, the distance to Tycho, we find that the required
ionisation fraction of the intervening ISM is $\chi_\mathrm{e} = \frac{\sqrt{EM \, l}}{N_\mathrm{H}} \sim 0.015 \sqrt{\frac{l}{2.5\,\mathrm{kpc}}}$
for $T=10$~K, or alternatively, $\chi_\mathrm{e} \sim 1.35 \sqrt{\frac{l}{2.5\,\mathrm{kpc}}}$ for $T=10,000$~K. 
The 10,000~K assumption for the diffuse ISM gas is more reasonable than the 10~K \citep{draine11}, although, of course, this gas does not extend
evenly along the line-of-sight to Tycho, but is likely in a patchy distribution (which would lower $\chi_\mathrm{e}$ to a more
reasonable value). We do not know the relative depths of these warm ionised
gas along the line-of-sight to Tycho, so unfortunately we cannot constrain a $\chi_\mathrm{e}$ for the case of this ISM.

Another point to note is that $\tau_{40, \mathrm{ISM}}=0.65$ corresponds to
an optical depth of $\tau_{30.9}=1.2$
at 30.9~MHz, although this is for a very small area (3.6~arcmin$^2$). \cite{kassim89} studied optical depths towards 15 Galactic SNRs,
and found only once source with  $\tau_{30.9}>1$. 
The integrated radio spectrum of Tycho (Fig. \ref{fig:radio_spectrum}) shows no indication of free-free absorption from the ISM kicking in at frequencies
lower than 100 MHz. There is a slight drop visible in the spectrum from LOFAR narrow band maps (Fig. \ref{fig:lofar_spectrum}), although this relies
only on the data point at 40~MHz.
For the integrated spectrum of Tycho's SNR we measure a best-fit $\tau_{30.9}=0.1$, well on the low side of the values measured by \cite{kassim89}.

The relatively high value of the optical depth in the region in green in Fig. \ref{fig:hba_reg}, and its small area suggest that
this is a small clump of ionised material. We cannot know if the clump is relatively close to the source or somewhere along the line-of-sight. 

Finally, the low-frequency absorption is only seen in a ring-like structure in the rim of the SNR and in two clumpy regions in the SNR
centre. In the remaining regions in the interior we do not find any detectable absorption. It is unlikely, though, that the foreground
ISM gas has the shape we see over Tycho, with a clear ring and a mostly empty interior. The regular morphology seen in the maps in
Fig. \ref{fig:Ems} does not favour the ionised ISM scenario as the dominant source of absorption.

\vspace{0.2cm}
\noindent
\textbf{2. Ionised diffuse cavity surrounding Tycho}

Consider that it is the cavity surrounding Tycho that is responsible for the ionisation we see at LOFAR frequencies. 

The size of the ionised cavity may influence the distributions of the foreground absorption.
As shown in figure \ref{fig:cartoon}, the depth of the ionised materials $l'$ is as a function of the projection radius $r$ ($r=0$ at the SNR center, $r=R$ at the SNR boundary), 
the radius of the SNR $R$ and the thickness of the cavity $l$ ($l' = \sqrt{l^2+2lR}$), resulting in:
\begin{equation}
  l'(R)\approx \begin{cases}
    \sqrt{2Rl}, & \text{if $l \ll R$ }\\
    l, & \text{if $l \gg R$}
  \end{cases} 
    \end{equation}
 \begin{equation}
  l'(0)=l.
      \end{equation}
If the cavity size is much larger than the SNR radius, we would see a uniform ionisation distribution as $l'(r)=l$. The ring-like ionisation distribution suggests that the cavity
is small and might be close to the SNR radius. 

\cite{williams13} found that 
the ISM density around Tycho is only $n_\mathrm{H}=0.1-0.2$~cm$^{-3}$, and that there is dust with temperature $T=100$~K. 

The optical depth value we report for the rim of Tycho (the region in red in Fig. \ref{fig:hba_reg}), $\tau_{40,\mathrm{ISM}}=0.29$, assuming $Z=1$ and
$T=100$~K, corresponds to an emission measure of
$EM=2.1$~pc~cm$^{-6}=n_\mathrm{e}^2 \,l_\mathrm{cav}$, where $l_\mathrm{cav}$ is the size of the cavity. 
This implies $n_\mathrm{e}=1.5\sqrt{\frac{l_\mathrm{cav}}{1~\mathrm{pc}}}~\mathrm{cm}^{-3}$. 
Recall that $n_\mathrm{e} = \chi_\mathrm{e} n_\mathrm{H}$.

\cite{woods17} measured the ionisation fraction of the ambient hydrogen ahead of the forward shock to be
$ \chi_\mathrm{e} < 0.2$ (the ambient hydrogen is more than 80\% neutral). 
They obtained the ionisation fraction for the atomic gas, which has a higher density; they used $ n_\mathrm{H}=1$~cm$^{-3}$.
Setting $\chi_\mathrm{e} = 0.2 = \frac{n_\mathrm{e}}{ n_\mathrm{H}}$
means that the cavity must be very small, $l_\mathrm{cav}<0.02$~pc. 

As mentioned above, a thin length for $l_\mathrm{cav}$ is supported by the geometry of the external absorption map, which appears to be limb-brightened.
However, this is a very restrictive value, requiring that Tycho be almost but not quite interacting with the
molecular cloud, and not just in one place but around its entire perimeter. This is very unlikely. 

\vspace{0.2cm}
\noindent
\textbf{3. Ionised dense molecular environment surrounding Tycho}

In this section we consider whether the ionised structure
is related to the
molecular cloud found by \cite{lee04} and discussed in \cite{zhou16}. The morphological coincidence of the molecular cloud in the north-east with
the region of highest absorption is suggestive of such a relation. 

\cite{zhou16} tabulate the molecular hydrogen column density $N_{\mathrm{H}_2}$
for several positions 
and find values around $7\times10^{20}$~cm$^{-2}$ in the area where we measure $\tau_{40,\mathrm{ISM}}=0.65$,
implying
 $EM=0.30$~pc~cm$^{-6}$ (here the conditions
$Z=1$, $T=10$~K do apply). Since $EM = n_\mathrm{e}^2 l$,
$N_\mathrm{H_2} = n_\mathrm{H_2} l$, and $\chi_\mathrm{e} = \frac{n_\mathrm{e}}{n_\mathrm{H_2}}$, the value 
$\frac{EM}{N_{\mathrm{H}_2}} = \chi_\mathrm{e} n_\mathrm{e} = 4.3\times10^{-4}~\mathrm{cm}^{-3}$ is independent of
the  size of the molecular cloud.

If we take the size of the molecular clouds to be of the order of Tycho \cite[ $l_\mathrm{MC}\sim5$~pc , see Fig. 1, bottom-right in][]{zhou16},
then $n_\mathrm{e} = 0.25 \sqrt{\frac{l_\mathrm{MC}}{5~\mathrm{pc}}}~\mathrm{cm}^{-3}$, which corresponds to 
$\chi_\mathrm{e} = 2\times10^{-3} \left( \frac{l_\mathrm{MC}}{5~\mathrm{pc}} \right)^{-1/2}$. 

Generally, dense molecular cores have $\chi_\mathrm{e} \sim 10^{-8} - 10^{-6}$ \citep{caselli98}, while translucent and diffuse 
molecular gas has typical $\chi_\mathrm{e} \lesssim 10^{-4}$ \cite[][figure 1]{snow06}. $\chi_\mathrm{e} \sim 10^{-3}$ requires
an external ionising source.

\vspace{0.5cm}
\noindent
It is not possible to tell directly from our observations of free-free absorption whether the ionised absorbing component
is in the environs of Tycho or far in the ISM along the line-of-sight. 
However, the fact that the absorption occurs where the remnant is
brighter and expanding into a higher density region \citep{reynoso99b, williams13} is suggestive to us of a local effect,
as is the rimmed geometry.
If the thin cavity surrounding Tycho and separating the SNR shock from the molecular ring were responsible
for the absorption, then the cavity would have to be very thin but at the same time the shock could not have reached the
molecular material \textit{anywhere} along its boundary ---a contrived geometry.
The high neutrals
values inferred by \cite{woods17}, the clear presence of Balmer shocks \citep{ghavamian00}, and the morphological coincidence
with the molecular cloud in the north-east all point towards the molecular material being associated with the absorption.
Finally, the bubble-like distribution of the molecular gas provides a natural explanation for the
rimmed absorption morphology. We conclude that the absorption is most likely
due to the presence of over-ionised molecular clouds.

\subsection{What mechanism is responsible for the ionisation of Tycho's surroundings?}
\label{sec:source}

A SIMBAD query towards the direction of Tycho gives no OB associations or bright stars that could be responsible for the
observed ionisation:
Tycho itself is the only likely ionising source towards this line-of-sight. The sources of ionisation could be the X-ray emission from
Tycho, the cosmic rays accelerated in the SNR, or perhaps the ionising radiation emitted by the supernova progenitor or the event
itself. A full discussion of the different ionisation scenarios requires a detailed treatment of ionisation and recombination in the modelling, 
and is beyond the scope of this paper.

\subsection{Internal absorption and mass in the unshocked ejecta}
\label{ush_mass}

The amount of mass in ionised material internal to the SNR reverse shock is given by \cite[see][]{arias18a}:
\begin{equation}
M =  A S l^{1/2} m_\mathrm{p} \frac{1}{Z} \sqrt{EM},
\label{eq_mass}
\end{equation}
where $A$ is the mass number of the ions, $S$ is the area of the region for which we measure the absorption, $l$ is the depth of the absorbing material, $m_\mathrm{p}$ is the 
mass of the proton, $Z$ is the number of charges, and $EM$ is the emission measure.
Making certain assumptions about these values, one can derive a value for the mass in unshocked material from our measured
optical depth. 

The easiest parameter to estimate is the mass number of the ions $A$. Tycho is the result of a Type Ia explosion; out of the
$\sim1.4$~\msun\ of ejecta it produced, $0.5-0.8$~\msun\ is expected to be iron \citep{badenes06}. 
In a spectroscopic analysis of ASCA data \cite{hwang98} noted that iron is in fact the most recently ionised element,
and so it is likely to compose the bulk of the unshocked material. 
\cite{hayato10} also found segregation of Fe in the inner ejecta from a study of the expansion velocities of the X-ray emitting material.
Moreover, the X-ray emission from iron in Tycho 
is not as prominent as in other type Ia SNRs \cite[e.g. Kepler, ][]{reynolds07}, suggesting that some of it is not visible in the X-rays yet.
For these reasons we take $A=56$, corresponding to Fe. We take $Z=3$, for three-times ionised Fe.

$S$ is the surface area of the absorbing region (the area in yellow in Fig. \ref{fig:hba_reg}). We do not know the thickness
of the absorbing slab $l$, which is actually critical for the mass determination, because we do not have a way of probing the
three-dimensional structure of the absorbing material. For a homogeneous distribution of material within the sphere of the reverse
shock, the average depth is $l=\frac{4}{3}R$ \cite[where $R$, the radius of the reverse shock, is 2.25~pc for a distance of 2.5~kpc, ][]{tian11}.

Finally, the value of the $EM$ depends on $Z$ and the temperature $T$. We do not know what the temperature conditions in 
the unshocked ejecta of Tycho are; an accurate determination would require infrared observations that could measure the ratios between
different forbidden lines of the ionised material. To our knowledge, the only time the temperature from the unshocked ejecta of a 
SNR has been measured is in the case of Cas A, whose unshocked ejecta has a temperature of 100~K \citep{raymond18}. 
Although it is not clear that the radiation from Tycho's SNR could maintain its internal material heated to 100~K, we will take this to
be the value in our mass estimate.

The $EM$ values in Table \ref{tb:fits_int} correspond to the following mass estimates:
\begin{equation}
\begin{split}
M = & 6.5 \pm 2.1 \, M_{\odot}\, \left(\frac{A}{56}\right) \left(\frac{l}{3.0 \,\rm{pc}}\right)^{1/2} \left(\frac{Z}{3}\right)^{-3/2} \\
& \left(\frac{T}{100~\mathrm{K}}\right)^{3/4}  \times \sqrt{\frac{g_{\mathrm{ff}}(T=100 \, \mathrm{K},Z=3)}{g_{\mathrm{ff}}(T,Z)}},
\end{split}
\label{eqn_mass}
\end{equation}
in the case of the upper limit with $EM=0.33$~pc~cm$^{-6}$, and in the case of $EM=179$~pc~cm$^{-6}$, $M=146 \pm 39$~\msun,
with the same parametrisation.

\subsection{What are the conditions and structure of the ejecta internal to Tycho's reverse shock?}

Our upper limit above is not useful, and 
the mass estimate for the $\alpha=0.71$ fit is completely unreasonable, since the total amount of ejecta resulting from the
explosion of Tycho's progenitor was $\sim1.4$~\msun. As we mention above, a determination of the $EM$ depends very much on the
expected flux if no absorption were present, but if there is indeed absorption noticeable at LOFAR HBA frequencies ($\sim150$~MHz), then the
high mass estimate value implies that
the conditions we assumed in the section above
do not describe the actual physical conditions internal to the SNR reverse shock.

Lowering the temperature or invoking a higher ionisation state alone are not sufficient to
arrive at a meaningful mass estimate. A further way to reduce the mass estimate for a given $EM_\mathrm{int}$ is 
if not all unshocked material is iron, but
lighter elements are also present. \cite{decourchelle17} notes that the comparison of iron-L complex and Si-K line images indicates good
mixing of the Si and Fe layers synthesised in the supernova. The mass number of Si is half of that of Fe, so if silicon is 
present, the mass estimate could be significantly reduced. 

The effects of temperature, ionisation conditions, and composition can be important if combined, but the single effect that 
can have the largest contribution to the high absorption value is the degree of clumping in the unshocked material.
The estimate in equation \ref{eqn_mass} assumes that the ejecta are distributed homogeneously within the sphere of the reverse shock.
This is what one expects for an ejecta density profile with a flat core and an exponential outer region \citep{chevalier82}, if the reverse shock has already reached
the core. 

\cite{sato19} analysed \textit{Chandra} observations of Tycho and found from its genus statistic that Tycho's X-ray ejecta structure
strongly indicates a skewed non-Gaussian distribution of the ejecta clumps, possibly from initially clumped ejecta. 
The radioactive decay of elements synthesised in the explosion could also cause the ejecta to have a foamy distribution, 
as is the case for Cas A \citep{milisavljevic15}.
If the unshocked ejecta in Tycho are heavily clumped it can be possible to see absorption in the LOFAR HBA even for modest amounts of
unshocked mass.

\section{Conclusions}

In this work we have mapped Tycho's SNR with the LOFAR Low-Band and High-Band Antennae, centred at 58~MHz and 143~MHz, 
respectively. These are the lowest-frequency resolved observations of this source to date, even though the angular resolution of our LBA maps
is modest (41\arcsec). We compared these maps to higher frequency VLA observations at 330~MHz and 1400~MHz \citep{katz-stone00, williams16},
and found that in some regions the LOFAR flux is  lower than expected for an
unabsorbed synchrotron source. 
We identify this effect as low-frequency free-free absorption due to foreground free electrons
absorbing the background synchrotron radiation from Tycho. 

It is unlikely, from the observed geometry, that the low-frequency absorption is due to line-of-sight material far away from Tycho,
but rather it must be in the environment of the SNR.
There are two regions that could be responsible for the ionisation: the diffuse, infrared-emitting region immediately
surrounding Tycho, or its neighbouring molecular clouds. If the former is true, and the absorption is due to an ionised cavity
surrounding Tycho, then this cavity must be very thin ($<0.02$~pc), so as to not contradict earlier results on the neutral fraction ahead of the
shock. Alternatively, if the molecular clouds are responsible for the absorption, then the implied ionisation fraction requires an
external ionising source. Tycho itself is the only candidate, through its X-ray emission, its cosmic rays, or possibly from the ionising
flux of its progenitor white dwarf or the supernova explosion.

Finally, we tried to measure the free-free absorption
in the region internal to the SNR reverse shock from its unshocked ejecta. However, we are limited by our knowledge of the
unabsorbed spectral behaviour of the source at these frequencies: the amount of absorption we measure depends on what is the spectral 
index in the region, which is poorly constrained due to systematics error and an incomplete knowledge of the spectral behaviour at high frequencies. 
According to our best-fit scenario, the spectral index in the region internal to the reverse shock is relatively
high and a copious amount of free-free absorption is required to explain the LOFAR flux densities.
If real, we attribute the absorption to cold, ionised, unshocked stellar ejecta inside the SNR reverse shock free-free absorbing the synchrotron
emission from the back side of the shell. In order to account for the high value of internal absorption we measure we expect the ejecta
to be colder than 100~K, be somewhat highly ionised, and be heavily clumped. 

Radio observations in the few GHz range could determine the unabsorbed, resolved spectral index of the source, and observations in the
$200-1000$~MHz range would allow us to better model the parameters responsible for the absorption, which result in a characteristic
spectrum with curvature at these frequencies. Finally, hyperfine structure infrared line observations of these clumps would be necessary 
to better understand their temperature and composition, both critical in determining the mass in unshocked ejecta.

\acknowledgments

We thank N. Kassim for the 330~MHz VLA image, and B. Williams for the 1.4~MHz VLA image. 

This paper is based (in part) on data obtained with the International LOFAR Telescope (ILT) under project code LC10\_011. LOFAR \citep{vanhaarlem13} is the Low Frequency Array designed and constructed by ASTRON. It has observing, data processing, and data storage facilities in several countries, that are owned by various parties (each with their own funding sources), and that are collectively operated by the ILT foundation under a joint scientific policy. The ILT resources have benefitted from the following recent major funding sources: CNRS-INSU, Observatoire de Paris and Universit\'e d'Orl\'eans, France; BMBF, MIWF-NRW, MPG, Germany; Science Foundation Ireland (SFI), Department of Business, Enterprise and Innovation (DBEI), Ireland; NWO, The Netherlands; The Science and Technology Facilities Council, UK.

We acknowledge the use of archival data from the National Radio Astronomy Observatory's Karl G. Jansky Very Large Array (VLA). The National Radio Astronomy Observatory is a facility of the National Science Foundation operated under cooperative agreement by Associated Universities, Inc.

\software{LOFAR Low-Frequency Pipeline \citep{degasperin19}, wsclean \citep{offringa14}, Pre-Facet Calibration Pipeline \citep{vanweeren16}, ddf-pipeline
\cite[v2.2;][]{shimwell19}, LMFIT: Non-Linear Least-Square Minimization and Curve-Fitting for Python \citep{newville14}, APLpy: Astronomical Plotting Library in Python \citep{robitaille12}.}

%

\vspace{5mm}
\facilities{The LOw Frequency ARray (LOFAR), the Karl G. Jansky Very Large Array (VLA).}





\bibliography{./october}



\end{document}